\documentclass[pra,reprint,superscriptaddress,amsmath,amssymb,aps,floatfix]{revtex4-1}

\usepackage{hyperref}
\usepackage{graphicx}
\usepackage{comment}

\usepackage{color}
\definecolor{purple}{rgb}{0.63,0,1}
\definecolor{dark-green}{rgb}{0,0.4,0.1}
\definecolor{dark-gray}{rgb}{0.4,0.4,0.4}

\definecolor{pink}{rgb}{1,0,0.9}

\newcommand{\change}[1]{{#1}}

\usepackage[notext]{stix}
\usepackage{times}

\DeclareSymbolFont{CMAlt}{OMX}{cmex}{m}{n}
\DeclareMathSymbol{\sumop}{\mathop}{CMAlt}{"50}

\usepackage{color}
\hypersetup{%
    unicode=true,
    colorlinks=true,
    linkcolor=blue,
    citecolor=blue,
    urlcolor=blue,
    pdftitle={Aspects of Bose-Einstein condensation in a charged boson system over the dielectric surface},
    pdfauthor={I.V. Lukin, A.G. Sotnikov, Yu.V. Slyusarenko},
    pdfproducer={},
    pdfcreator={}}

\newcommand{\brho}{\boldsymbol{\rho}}	
\newcommand{\sgn}{\operatorname{sgn}}

\newcommand{\subfigref}[2]{\hyperref[fig:#1]{\ref*{fig:#1}(#2)}}
\newcommand{\subfigrefs}[3]{\hyperref[fig:#1]{\ref*{fig:#1}(#2)--(#3)}}

\begin{document}
\title{Aspects of Bose-Einstein condensation in a charged boson system over the dielectric surface}

\author{I.V. Lukin}
\email{lukin.ilya@yahoo.com}
\affiliation{Karazin Kharkiv National University, Svobody Sq. 4, 61022 Kharkiv, Ukraine}

\author{A.G. Sotnikov}
\author{Yu.V. Slyusarenko}
\affiliation{Karazin Kharkiv National University, Svobody Sq. 4, 61022 Kharkiv, Ukraine}
\affiliation{Akhiezer Institute for Theoretical Physics, NSC KIPT, Akademichna 1, 61108 Kharkiv, Ukraine}

\date{\today}

\begin{abstract}
We study theoretically a gas consisting of charged bosons (ions) over the flat dielectric surface at low temperatures and its tendency to form a state with a Bose-Einstein condensate.
For the stability of a system, an additional external electric field, which keeps charges at the dielectric surface, is introduced.
The formalism is developed in the framework of a self-consistent-field approach, which combines the quasiclassical description in terms of the Wigner distribution functions and the quantum-mechanical approach by employing the Gross-Pitaevskii equation.
We predict a formation of the state with a Bose-Einstein condensate and determine the near-critical physical characteristics of the system.
It is shown that the thermal and condensate components become spatially separated under these conditions.
{We discuss the limitations of the developed semiclassical approach and prospects for the pure quantum-mechanical treatment of the problem.}
\end{abstract}

\maketitle

\section{Introduction}
The studies dedicated to the effects and phenomena in various systems with a Bose-Einstein condensate (BEC) rely on almost a century of history~\cite{Bos1924ZP, Ein1925SKPAW}.
The condensation phenomenon, in particular, is responsible for superfluidity and superconductivity, but these have been successfully explained within phenomenological theories as well. As for unambiguous evidence, it took around 70 years to observe it directly in dilute gases~\cite{Cor1995S, Ket1995PRL, Bradley1995PRL}.
The reason consists in difficulties in achieving quantum degeneracy regime in atomic gases, which requires involving the laser cooling techniques~\cite{Phillips1998RMP}. A more detailed overview of the history of research on the physics of BECs in ultracold gases can be found, e.g., in Refs.~\cite{Pethick2002, Pitaevskii2003}.

The proof of the existence of a BEC under experimental conditions has opened new directions in experimental and theoretical studies of quantum gases.
Among these, a special attention is given to the studies of the interaction of electromagnetic waves with ultracold gases in the presence of a BEC, in particular, the ultraslow-light phenomena~\cite{Hau1999N, Sly2008PRA}. These studies open up prospects for various practical applications, see, e.g., Refs.~\cite{Zhang2009PRL, Sly2009PLA, filter2010ltp, relat2011pra}.
In this regard, the studies related to the experimental implementation of a BEC of photons should be mentioned as well~\cite{Klaers2010, 10.1117/12.2001831}. The main feature of these experiments is the fact that condensation can be achieved only in the presence of a substance, a dye, in particular.
Due to this feature, the BEC of photons can be realized at room temperatures. Furthermore, the condensation of photons in ultracold gases opens up a possibility of a coexistence of condensates of atoms and photons \cite{PhysRevA.88.013615, Boichenko_2015}.
It should be noted that photons in a medium cease to be ``pure'' photons, being, in fact, quasiparticles with a photon component. Without this, their thermalization would be impossible with a change of the temperature of a substance, which is necessary for realization of the BEC of photons in certain modes. 
For this reason, it is worth mentioning the studies on a Bose-Einstein condensation of exciton polaritons. Interest to this research direction was initiated by Ref.~\cite{Kasprzak2006}, which was followed by a large number of publications on this topic.
A significant part of publications devoted to the condensation of polaritons is summarized in the overview~\cite{proukakis_snoke_littlewood_2017}, where characteristics of polariton condensates are compared to those of atomic condensates. 
There is also a number of other systems with unique properties, in particular, 
atomic Bose-Einstein condensates dressed by electromagnetic field giving rise to soliton physics \cite{Dong2013, Qin2015, Qin2016, Qin2019} and
condensates of heteronuclear molecules formed in two-component Fermi gases~\cite{Peletminskii_2017}.


\change{As for the research on systems of charged bosons, the central object of the study, the earliest interest to the BEC phenomenon in these systems appeared, apparently, in 1960's \cite{PhysRev.124.649, PhysRev.127.1809, 1963PhL.....4..278N, NINHAM1964685, PhysRev.143.91}.}
It was motivated mainly by possible applications in the physics of superfluidity, superconductivity, physics of normal metals, nuclear physics, and astrophysics.
However, the obtained results were viewed as inconclusive, 
primarily because of the limited applicability of
the Bogoliubov method~\cite{Bogolyubov:1947zz} for dense systems.
Among the succeeding theoretical studies, Refs.~\cite{PhysRevB.12.2619, Dolgov_2009, Dolgov2012, doi:10.1063/1.4979959} and references therein should be mentioned. These were still focused rather on dense systems, e.g., superfluids, which under certain conditions can be described as Coulomb systems~\cite{doi:10.1063/1.4979959}, plasma or even cosmological systems~\cite{Dolgov_2009, Dolgov2012}. A special attention was given to the effect of the BEC of charged bosons on the Debye screening in plasma (or in Coulomb systems), as well as on the propagation and polarization of photons in cosmological plasma.

In all problems related to the study of equilibrium states of systems of charged bosons, a question of compensation for the mutual repulsion of positively charged bosons inevitably arises.
In Refs.~\cite{Dolgov_2009, Dolgov2012, doi:10.1063/1.4979959,Sakaguchi2020} and references therein, the issue of compensation is solved by introducing oppositely charged particles.
For this reason, a special attention is given to studies of the effect of screening of charges and the influence of a BEC consisting of positive charges on it. In a number of studies, including Refs.~\cite{PhysRevB.12.2619, PhysRev.124.649, PhysRev.127.1809, 1963PhL.....4..278N, NINHAM1964685, PhysRev.143.91, PhysRevB.12.2619, Dolgov_2009,Dolgov2012, doi:10.1063/1.4979959}, the existence of compensation for the Coulomb repulsion is simply postulated.
It is necessary to stress that the above-mentioned studies are applied to the dense systems with, at least, two reasons for this preference.
The first one originates from the fact that the BEC phenomena were stimulated by potential applications in physics of superfluidity, superconductivity, normal metals, etc. 
The second reason, most likely, is that it was difficult to realize a real physical system (a gas) consisting of identical charged bosons with low density. 
Moreover, in such a system, as noted above, the mutual repulsion of positively charged bosons should be compensated and the BEC regime must still remain within the experimental reach. However, nowadays such a system can be realized due to the progress in the laser cooling and trapping techniques.

\change{In this paper, we study a system of charged bosons above the surface of solid dielectric as a new physical system with a possibility of condensation.
In this case, a dielectric also plays an essential role in creating the necessary conditions for achieving a BEC phase in the system. 
It enters as an external control parameter through specific boundary conditions for the charged bosons and the electric field created in the system.}
The idea arises from an intensive research related to phenomena and effects in systems of electrons above the surface of a dielectric, in particular, liquid helium, see, e.g., Ref.~\cite{Monarkha:1625049} and references therein. A particular interest in these problems stems from the so-called phenomenon of the Wigner crystallization of electrons in metals~\cite{PhysRev.46.1002, doi:10.1063/1.593743}.
Since there was no direct experimental confirmation of the existence of three-dimensional crystal structures predicted by Wigner, the research turned to electrons on the surface of liquid helium.
The experiments succeeded in realizing various two-dimensional periodic electronic structures above the surface of liquid helium~\cite{PhysRevLett.23.1238, 1971JETP...33..387S, PhysRevLett.29.1233, 1974PhRvL..32..280G}.
Theoretical approaches to the description of these two-dimensional spatially periodic structures are often based on the concept of the so-called levitating electrons~\cite{Monarkha:1625049}.
The key idea of this approach is a description of a single charge above the surface of the dielectric, which, together with the electrostatic image in the latter, can be interpreted as a one-dimensional hydrogen-like system with the corresponding energy spectrum.
The hydrogen-like structure of the spectrum leads to the localization of such a quantum-mechanical object in the ground state at a certain distance from the dielectric surface.
In most cases, this allows to neglect the influence of the surface inhomogeneities on the state of a charge.
But the gas of charges at the surface of the dielectric is a many-body system.
For this reason, due to a long range of the Coulomb forces, the correctness of considering charges as single non-interacting particles at the dielectric surface becomes doubtful even in the low-density limit. 
To solve this problem, a so-called ``palliative'' approach was developed.
Namely, it was stated that the charge is placed in an external model potential.
This potential is a Coulomb potential modified by means of two tunable constants.
These constants are related to the final value of the surface potential barrier and the final value of the potential on the dielectric surface~\cite{Monarkha:1625049}.

It is also clear that the electrostatic potential in the system under consideration is a quantity consistent with the distribution of charges over the surface of the dielectric. This represents a self-consistent field (SCF). 
Hence, in a consistent theory without tunable parameters, this potential must be determined simultaneously with the distribution function of charged particles over the dielectric surface. One of methods of the SCF theory suitable for solving this problem is the Thomas-Fermi approach~\cite{2009arXiv0905.3294L, doi:10.1063/1.4753978, Lytvynenko_2017, 2018arXiv180909921L}. 
When applied to many-electron atoms, this method is based on the Poisson equation, which relates the electrostatic potential to the distribution of electrons in an atom \cite{Landau5eng}.
Important in this model is also the concept of a positively charged nucleus of an atom, which compensates the Coulomb repulsive forces between electrons.
In the problem of the distribution of charges and fields in a system of charges over the surface of a dielectric, the factors compensating repulsive forces between particles are the polarization forces and the external attractive electric field.
The latter, which attracts charges to the surface and affects their spatial distribution, is used in experiments to observe the Wigner crystallization over the surface of liquid helium~\cite{Monarkha:1625049}.

A potential of the approach developed in Refs.~\cite{2009arXiv0905.3294L, doi:10.1063/1.4753978, Lytvynenko_2017, 2018arXiv180909921L} to the study of the many-body states of electrons above the helium surface is evident.
The microscopic theory of charged fermions in an external electric field above the surface of a dielectric is based on the first principles of quantum statistical mechanics. 
The self-consistent equations for the charge distribution function, the electric field potential, and the shape of the surface of the liquid dielectric can be obtained.
These equations are employed to analyze the phase states of the system under study and phase transitions, in particular, to the states with spatially-periodic ordering.
In the next section, this theory will be generalized to the system of charged bosons including a phase transition to the BEC state.

\section{Formalism}\label{sec:formalism}

\subsection{Equations of electrostatics}\label{sec:El_eqs}
\begin{figure}
  \includegraphics[width=\linewidth]{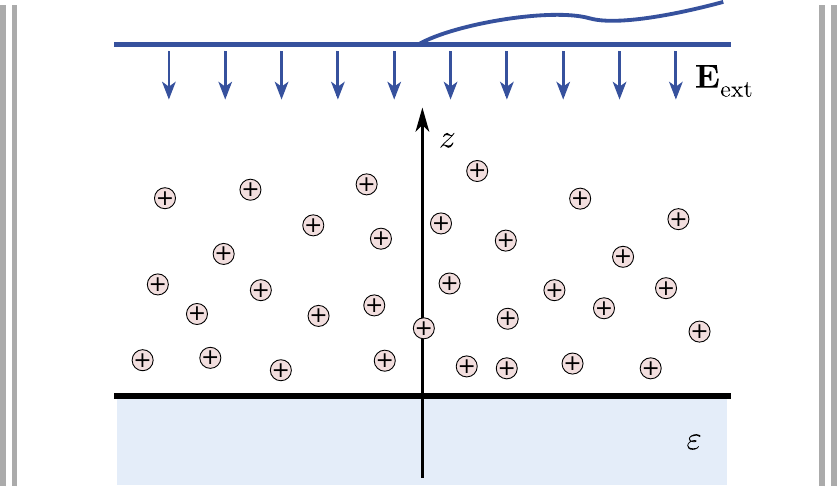}
  \caption{\label{fig:schematic}%
    Schematic representation of the system under study.}
\end{figure}

We describe a system consisting of $N$ particles (indistinguishable bosons), each having the charge~$q$ and the integer spin~$s$.
{Although charge carriers (ions) can have a complex charge distribution, for simplicity, we account only for the leading monopole terms in the corresponding multipole expansion of the interparticle potential.}
These particles are placed in a vacuum above the flat surface of a medium characterized by the homogeneous dielectric constant~$\varepsilon$,
as schematically illustrated in Fig.~\ref{fig:schematic}.
We assume the dielectric is not limited from below and non penetrable for the charge carriers. We also assume that the attractive electric field $E_{\rm ext}$ 
acts along the $z$ axis toward the interface at $z = 0$. It should be mentioned that forces of dielectric polarization attract the free charges to the surface. 
However, in the absence of an external ``restraining'' field, the polarization forces on their own are insufficient to stabilize the many-particle system, which will be discussed further below. To avoid the problem of particle repulsion in the plane of interface, we assume that the system is placed in a vessel with dielectric non penetrable walls at $|\brho|\to\infty$, where $\brho = (x,y)$ is a radius-vector in the corresponding symmetry plane.
These walls prevent charges from leaving the system.

Within this paper, we do not discuss in detail the methods to prepare a system of identical charged bosons. It is possible, for example, to imagine a system of identical positively charged ions with integer spin immersed into a vessel with impenetrable dielectric walls. 
In the case under study it is sufficient to assume that the system geometry allows the physical conditions to be realized. Namely, we assume that ions are mainly concentrated above one of the walls of the vessel, which can be viewed as flat, and that the influence of other walls can be neglected. This assumption can be the first step in the theoretical description as it substantially simplifies the analysis. The proposed theory allows generalization to systems with more complex geometry, however, analytical results in more complex geometries are unavailable. Some methods to cool down and accumulate ions can be found, e.g., in Ref.~\cite{Tomza2019RMP} and references therein.

Let us now turn to the derivation of the SCF equations for the system under consideration. 
As it is mentioned, one can start immediately from the Poisson equation, where the charge density is determined in a self-consistent way by the sum of potentials produced by the external field, charges themselves and their respective image charges, see Ref.~\cite{2009arXiv0905.3294L}. However, in this case we have certain ambiguities in the strength of electric field either at $z=0$ or at infinity. These problems are connected with a quasi-one-dimensionality  of the system. We can avoid them by deriving the one-dimensional Poisson equation from the potential of a single charge above the dielectric surface and averaging it with the corresponding distribution function.

To this end, let us specify the potential produced by a single particle (boson), which is located above the flat surface of a dielectric with the permittivity~$\varepsilon$.
Denoting its charge and position by $q$ and ${\bf r}_0\equiv (x_0,y_0,z_0)\equiv (\brho_0,z_0)$, respectively, the corresponding electrostatic potential at the point~${\bf r}$ can be written as
\begin{equation}\label{eq:phi_single}
  \varphi^{(0)}({\bf r})
  =\frac{q}{|{\bf r}-{\bf r}_0|}
  -
  \frac{q'}{|{\bf r}-{\bf r}_0'|},
\end{equation}
where ${\bf r}_0'\equiv (\brho_0,-z_0)$ and $q'\equiv q ({\varepsilon-1})/({\varepsilon+1})$ are the position and the absolute value of the corresponding image charge, respectively.
In case there is a large number $N$ of charges at $z>0$, these can be approximated by a continuous density~$n({\bf r}_0)$. 
{ Furthermore, we omit any specific in-plane arrangement of the charges meaning that the gas is dense enough for the Wigner crystallization instability to be suppressed (thus, the mean interparticle distance must be smaller than 10~$\mu$m).}
Therefore, the total potential produced by this system takes the form
\begin{equation}\label{eq:phi_total}
  \varphi({\bf r})
  =q\int\limits_{z_0>0}d{\bf r}_0 n({\bf r}_0)
  \left(
    \frac{1}{|{\bf r}-{\bf r}_0|}
    -\frac{\varepsilon-1}{\varepsilon+1}\frac{1}{|{\bf r}-{\bf r}_0'|}
  \right).
\end{equation}
Note that the integral in Eq.~\eqref{eq:phi_total} is taken over the whole system volume above the dielectric.
In accordance with the assumptions made above, we consider a quasi-one-dimensional problem, where the plane of the dielectric surface is infinite in the $x$ and $y$ directions, i.e., the system can be viewed as axially symmetric in this plane, thus
\begin{equation}\label{onedim}
    n({\bf r}_0)=n(z_0).
\end{equation}

Due to the symmetry of the problem, the electric field components $E_{x}$ and $E_{y}$ vanish.
In turn, for $E_{z}$ by means of Eqs.~\eqref{eq:phi_total} and  \eqref{onedim}, we obtain the following expression:
\begin{eqnarray}
  E_z
  &=&
  2\pi
  q\int\limits_{0}^\infty
  dz_0 n(z_0)
  \int\limits_{0}^\infty
  {\rho}d{\rho}
  \left\{
    \frac{z-z_0}{
    \left[
    {\rho}^2
    +(z-z_0)^2\right]^{3/2}}
  \right.
  \nonumber
\\
  &&\qquad
  \left.
    -\frac{\varepsilon-1}{\varepsilon+1}
    \frac{z+z_0}{
    \left[
    {\rho}^2
    +(z+z_0)^2\right]^{3/2}}
  \right\}.
  \label{eq:Ez_1}
\end{eqnarray}
The integral is calculated in the cylindrical coordinate system and we obtain the following result:
\begin{equation}\label{eq:Ez_2}
  E_z = 
  2\pi
  q\int\limits_{0}^\infty
  dz_0 n(z_0)
  \left[
   \sgn(z-z_0) -\frac{\varepsilon-1}{\varepsilon+1}\right],
\end{equation}
where we used the property that $z$ and $z_0$ are positive.
This equation can be also transformed into the form:
\begin{equation}\label{eq:Ez_3}
\begin{split}
    E_{z}(z) &= 2\pi q\int_{0}^{z} n(z_0)dz_0 
    -2\pi q\int_{z}^{\infty} n(z_0)dz_0 \\
    &-2\pi q'
    \int_{0}^{\infty}n(z_0)dz_0.
\end{split}
\end{equation}
From the last term it is evident that the polarized dielectric generates a constant electric field, which attracts free charges to the surface. The expression~\eqref{eq:Ez_3} allows obtaining the boundary conditions for the electric field at $z=0$ and $z=\infty$,
\begin{equation}\label{boundE}
    E_{z}(0)= -\frac{4\pi qN_S\varepsilon}{\varepsilon+1} < 0,
    \quad
    E_{z}(\infty) = \frac{4\pi qN_S}{\varepsilon+1} > 0.
\end{equation}
Here, the quantity~$N$ denotes the number of charged particles per unit area~$S$ in the symmetry plane,
\begin{equation}\label{ns}
   N_S\equiv{N}/{S}=\int_{0}^{\infty}n(z)dz.
\end{equation}
Indeed, the total number of particles ${N}$ is determined in terms of the density $n({\bf r}_0)$ by ${N}= \int d{\bf r}_0 n({\bf r}_0)$.

From the boundary conditions~\eqref{boundE}, it is clear that at $z=0$ the electric field is negative, so it is directed to the surface of dielectric. At infinity, the field is, vice versa, positive, excluding the case $\varepsilon = \infty$, when the dielectric completely ``screens'' the field of original free charges resulting in $E_{z}(\infty)=0$. In a general case, therefore, there is a point ${z}^*$ at which the electric field vanishes. Presence of such a point is a sign of instability. Indeed, if one  places a probe positive charge in the region $z<{z}^*$, it will be attracted to the surface of dielectric. 
On the other hand, the charge placed at $z>{z}^*$ experiences a repulsive force acting in the direction $z\to\infty$. 
In other words, under these conditions the gas of charged particles will be unstable. 
This yields the necessity of introducing an external ``restraining'' field~${\bf E}_{\rm ext}$, homogeneous in the upper half-space and directed to the dielectric surface, see Fig.~\ref{fig:schematic}.  
Below, we denote by $E_{\rm ext}$ its absolute value, so that $E_{\rm ext} \geq 0$. Given the presence of this external field, Eq.~\eqref{boundE} can be transformed to the following form:
\begin{equation}\label{boundE2}
    E_{z}(0)= -\frac{4\pi qN_S\varepsilon}{\varepsilon+1}-E_{\rm ext},
    \quad
    E_{z}(\infty) = \frac{4\pi qN_S}{\varepsilon+1}-E_{\rm ext}.
\end{equation}

For stability of the system, as it is clear from the discussion above, it is necessary for the field to be constrained by inequality $E_{z}(\infty) \leq 0$. In other words, the system is stable if the following inequality is valid:
\begin{equation}\label{E'}
    E_{\rm ext} \geq \frac{4\pi qN_S}{\varepsilon+1}.
\end{equation}
Note here that the equality in Eq.~\eqref{E'}, which corresponds to the limit of stability of the system, corresponds also to the zero value of the total field at infinity, 
$E_{z}(\infty) 
=0$. For this reason, the system, where the equality in Eq.~\eqref{E'} holds exactly, can be called \emph{quasineutral}, since the field produced by the charges is completely compensated. Away from this condition, the system is ``charged'', since the field at infinity is not exactly compensated (in this regard, see also Refs.~\cite{Monarkha:1625049, PhysRevLett.23.1238, 1971JETP...33..387S, PhysRevLett.29.1233, 1974PhRvL..32..280G, 2009arXiv0905.3294L, doi:10.1063/1.4753978, Lytvynenko_2017, 2018arXiv180909921L}).
Thus, the validity of Eqs.~\eqref{boundE2} and \eqref{E'} guarantees stability of the system. Any equality from Eq.~\eqref{boundE2} can be used as a boundary condition for the first derivative of the potential~$\varphi(z)$.

The relation determining the potential $\varphi(z)$ can be derived by taking derivative of the equation
\begin{equation}\label{eq:pot}
\begin{split}
    \frac{d\varphi(z)}{dz} &= -2\pi q\int_{0}^{z} n(z_0)dz_0 
    +2\pi q\int_{z}^{\infty} n(z_0)dz_0 \\
    &+2\pi q'
    \int_{0}^{\infty}n(z_0)dz_0 + E_{\rm ext},
\end{split} 
\end{equation}
which is deduced from Eq.~\eqref{eq:Ez_3} with the external field $E_{\rm ext}$ and from the relation between the electric field and potential.
Finally, as expected, we arrive at the one-dimensional Poisson equation:
\begin{equation}\label{eq:Poisson1}
    \frac{d^2 \varphi(z)}{dz^2} = -4\pi qn(z),
\end{equation}
where the charge density~$n(z)$, in turn, is determined by the electrostatic potential created by both the charged particles themselves and by the external electric field. 
In other words, Eq.~\eqref{eq:Poisson1} plays a role of a SCF equation, which must be supplemented by equations determining the density of particles in terms of the potential~$\varphi(z)$.

\subsection{SCF approximation}
To formulate the closed self-consistent system of equations for the potential, we specify that the charged particles are bosons, i.e., their equilibrium state is described
in terms of the Bose-Einstein distribution function~$f (z, \bf p)$.
Taking into account the quasi-one-dimensionality of the problem,
\begin{equation}\label{distr_fun}
    f(z, {\bf p}) = \frac{1}{e^{\beta [\varepsilon ({\bf p})+q\varphi(z)-\mu]}-1},
\end{equation}
where $\varepsilon({\bf p})=p^2/2m$ is the dispersion relation of free nonrelativistic particles, $\mu$ is the chemical potential and $\beta$ is the inverse temperature in energy units, $\beta=1/T$ ($k_{\rm B}=1$). 

The total number of particles ${N}$ is obtained by integrating Eq.~\eqref{distr_fun} over the system volume~$V$ in the upper half-space and summing over all particle momenta,
\begin{equation}\label{Num}
\begin{split}
    {N}  &= \int d{V} \sum_{\bf p} f(z, {\bf p})
    =\frac{4\pi S\varsigma}{(2\pi\hbar)^{3}}
    \int dz \int _{0}^{\infty}dp p^2  
    \\
     &\qquad
    \times\left\{
        {\exp\left[
            {\beta ({{ p}^2}/{2m}+q\varphi(z)-\mu)}
        \right]
    -1}\right\}^{-1}.
\end{split}
\end{equation}
This equation determines the relation between the chemical potential, the gas temperature, and the number of particles in the system. In the last part of this formula, the summation was replaced by integration according to the standard procedure. The presence of the integer factor $\varsigma=(2s + 1)$ reflects the degeneracy of the single-particle states of bosons by their spin projection~$m_s=\{-s,\ldots,s\}$. 
Therefore, according to Eq.~\eqref{ns}, we determine the density of particles $n(z,T)$ as a function of the distance $z$ and the temperature~$T$ as 
\begin{equation}\label{density}
    n(z,T) = 
    \varsigma \lambda_T^{-3}
    g_{{3}/{2}}
    \left(e^{\beta\mu-\beta q\varphi(z)}\right),
\end{equation}
where $\lambda_T$ is the thermal de Broglie wavelength, $\lambda_T=(2\pi\hbar^2/mT)^{1/2}$, and
$g_{n}(x)$ is the so-called 
Bose-Einstein function \cite{Pethick2002, Pitaevskii2003}, 
$
    g_{n}(x)= 
    {\Gamma^{-1}(n)}
    \int_0^\infty 
    {y^{n-1}}{(x^{-1}e^{y}-1)^{-1}dy} .
$
This function is directly related to the known polylogarithm function, which is defined in the unit disk of the complex plane by the infinite series
$
    g_{n}(x) = \sum_{k=1}^{\infty} 
    {x^{k}}/{k^{n}}
$.
Thus, Eq.~\eqref{density} complements Eq.~\eqref{eq:Poisson1} transforming it into a closed equation for the SCF potential~$\varphi (z)$.

As it is evident from Eq.~\eqref{boundE2}, the electrostatic potential has a minimum at $z=0$. Using the fact that the potential is defined up to an additive constant, we set $\varphi(0)=0$. 
This particular choice for the boundary condition will further serve for a simplification of equations related to characteristics of a BEC including the corresponding transition temperature.
To summarize, according to Eqs.~\eqref{eq:Poisson1} and \eqref{density}, the state of a system consisting of the charged bosons is described by the Poisson equation of the type
\begin{equation}\label{Poisson}
    \frac{d^{2} \varphi(z)}{dz^{2}} = -4\pi q
    \varsigma \lambda_T^{-3}
    g_{{3}/{2}}\left(e^{\beta\mu-\beta q\varphi(z)}\right)
\end{equation}
with the boundary conditions for the potential $\varphi(0) = 0$, as well as for the electrostatic field given by Eq.~\eqref{boundE2}.
It is supplemented by the normalization condition
\begin{equation}\label{norm}
    \int_{0}^{\infty} 
    \varsigma \lambda_T^{-3}
    g_{{3}/{2}}\left(e^{\beta\mu-\beta q\varphi(z)}\right) dz = N_S,
\end{equation}
valid by virtue of Eqs.~\eqref{ns} and \eqref{density}. We would like to note that the boundary conditions~\eqref{boundE2} at $z=0$ and $z=\infty$ are equivalent, since the field is constrained by the 
relation $E_{z}(0)=E_{z}(\infty)-4\pi qN_S$.

It should be noted that Eq.~\eqref{Poisson} can be simplified
due to a specific form of its right-hand side, since the following recurrent relation holds for polylogarithms:
\begin{equation}\label{diff}
    \frac{d}{dx} g_{{5}/{2}}\left(e^{-x}\right) 
    = - g_{{3}/{2}}\left(e^{-x}\right).
\end{equation}
Using this fact, as well as the boundary conditions~\eqref{boundE2}, we reduce the order of the differential equation~\eqref{Poisson}.
To this end, according to the standard procedure, both sides of Eq.~\eqref{Poisson} are multiplied by $d\varphi/dz$ and integrated. Therefore, one arrives at the first-order differential equation for the electrostatic potential,
\begin{equation}\label{newreduced}
    \frac{d\varphi (z)}{dz} = \sqrt{8\pi 
    \varsigma T\lambda_T^{-3}
    g_{{5}/{2}}\left(e^{\beta\mu-\beta q\varphi(z)}\right)+E_{z}^{2}(\infty)}.
\end{equation}

Let us now turn to the normalization condition~\eqref{norm}. Using the same recurrent relation~\eqref{diff}, the integral in Eq.~\eqref{norm} can be evaluated analytically. Thus, we arrive at the relation determining the chemical potential in terms of the temperature, the strength of the external electric field, and the number of particles $N_{S}$, 
\begin{equation}\label{eq_of_state}
     \varsigma T\lambda_T^{-3}
    g_{{5}/{2}}\left(e^{\beta\mu}\right) 
    = qN_S\left(E_{\rm ext} +2\pi {q'}N_S
    \right).
\end{equation}
While calculating the integral in Eq.~\eqref{norm} to obtain~\eqref{eq_of_state}, it becomes necessary to change the integration variable. This replacement raises the question of the value of the potential $\varphi (z)$ at infinity. To address it, we note that the potential is a monotonically increasing function. In addition, the electric field is distributed in the way that it remains finite including $z=\infty$, see Eqs.~\eqref{boundE2} and \eqref{E'}. For this reason, the value of the potential at infinity should be taken infinite, $\varphi (\infty) = \infty$. 
The formulas~\eqref{Poisson}--\eqref{eq_of_state}  will further serve as the basis for analyzing the possibility of realizing the phase with a BEC in the system under study.

\subsection{Phase transition to the BEC state}\label{sec:BEC}

The critical temperature can be determined from Eq.~\eqref{eq_of_state} according to the condition
$T_c=T(\mu=0)$. Therefore, since $g_{5/2}(1)=\zeta(5/2)$, where $\zeta$ is the Riemann zeta function, it can be written as
\begin{equation}\label{eq:Tc-gen}
    T_c =\left[
    \frac{qN_S(E_{\rm ext} +2\pi {q'} N_S
    )   }
    {\varsigma\zeta(5/2)}
    \left( \frac{2\pi \hbar^2}{m} \right)^{3/2}
    \right]^{2/5}.
\end{equation}
For the quasineutral system with $E_{\rm ext}=4\pi qN/(\varepsilon+1)$ this yields
\begin{equation}\label{eq:Tc-qneutr}
    T_c =\left[
    \frac{2\pi q^2N_S^2}
    {\varsigma\zeta(5/2)}
    \left( \frac{2\pi \hbar^2}{m} \right)^{3/2}
    \right]^{2/5}.
\end{equation}
One can see that in this case the critical temperature has a power-law dependence on the number of particles per unit area, $T_c\propto N_S^{4/5}$.
Eq.~\eqref{eq:Tc-qneutr} allows to obtain estimates of characteristic temperatures for the given system parameters.

In particular, one finds that for a gas consisting of the singly-ionized $^{40}$K atoms with $\varsigma=9$ (the nuclear spin $I=4$) $T_c\approx0.2$~K and $T_c\approx1.3\times10^{-4}$~K at $N_S=10^{12}$~cm$^{-2}$ and $N_S=10^{8}$~cm$^{-2}$, respectively. In turn, in gases of ionized $^6$Li atoms the characteristic values of $T_c$ are approximately 5 times higher at the same densities due to the lower degeneracy $\varsigma=3$ (the nuclear spin $I=1$) and atomic mass~$m$.
Note that the chosen lower value $N_S=10^{8}$~cm$^{-2}$ corresponds to the mean distance $d\approx1~\mu$m between ions (under assumption of a quasi-2D arrangement, as we observe below), i.e., of the same order as realized in recent experiments on trapping and guiding cold ions, see, e.g., Refs.~\cite{Antohi2009,Romaszko2020}. 
The lowest value for $N_S$ can also be viewed as the lowest bound for the applicability of the SCF approach not only from the point of view of the Wigner crystallization at lower densities, but also due to the fact that in more dilute system the quantum fluctuations become more important thus invalidating the description by means of the Gross-Pitaevskii equation discussed below.

\subsection{SCF approximation in the presence of a BEC}\label{sec:formalism_below}
Below the critical temperature~$T_c$, the presence of a BEC becomes an essential factor, influencing physical characteristics of the gas of charged bosons above the dielectric surface. 
In the previous subsections, the formulation of the mean-field model did not include the presence of a condensate.
Meanwhile, the main statements of the proposed SCF theory require a significant modification in the presence of a BEC. 
The main reason is the necessity to modify the determination of the density~\eqref{density}, since the mean-field approximation in the form of Eqs.~\eqref{distr_fun}--\eqref{eq_of_state} is based on the semiclassical approximation for the Wigner function. Indeed, the Wigner functions use both the coordinate and momentum representations simultaneously. Thus, the quantum states, which are described by Eqs.~\eqref{distr_fun}--\eqref{Poisson}, must be occupied by a microscopic number of particles.
But the BEC itself corresponds to a coherent state of matter, where a macroscopically large number of particles is accumulated in the ground state. The outlined observations require a significant modification of the theory.

Below, we propose such a modification by ``hybridizing'' the SCF approximation. In particular, we assume that the density of particles above the dielectric surface is given by the sum of two terms, namely, the condensate and thermal components with the densities~$n_{0}$ and $n_{T}$, respectively.
\begin{equation}\label{eq: cond}
\begin{split}
    &n(z) = n_{0}(z)+n_{T}(z),\\
    &n_{0}(z) \equiv N_0(T) |\psi(z)|^{2},\\
    &n_{T}(z)\equiv {\varsigma}{\lambda_{T}^{-3}} g_{{3}/{2}}\left(e^{-\beta q\varphi(z)}\right).
\end{split}
\end{equation}
Note that the latter has the same form as Eq.~\eqref{density} with $\mu=0$ being the necessary condition for a phase transition to the BEC state, see Sec.~\ref{sec:BEC}. 
The first term in Eq.~\eqref{eq: cond} determines the condensate contribution to the total number of particles, where $N_0(T)$ is the number of the condensate particles per surface area~$S$. 
The quantity~$\psi(z)$ in Eq.~\eqref{eq: cond} is a wave function of a coherent state, i.e., a BEC. This function is also called as the Gross-Pitaevskii wave function that describes the normalized density of particles in the condensate,
\begin{equation}\label{eq:normwavefun}
    \int_{0}^{\infty}|\psi(z)|^2 dz = 1.
\end{equation}

Since the total number of particles per unit area must be preserved, $N_T(T)+N_0^{0}(T) = N_S$, according to Eq.~\eqref{norm} we can write
\begin{equation}\label{normcond2}
    N_0(T) = N_S - \int_{0}^{\infty}{\varsigma}{\lambda_{T}^{-3}} g_{{3}/{2}}\left(e^{-\beta q\varphi(z)}\right)dz.
\end{equation}
This is viewed as one of the SCF equations for obtaining $N_{0}(T)$. The wave function~$\psi (z)$ is, in turn, determined from the Gross-Pitaevskii equation, which can be expressed as
\begin{equation}\label{eq:grosspitaev}
    {\cal E}_0
    \psi(z)
    =-\frac{\hbar^2}{2m}\frac{d^2 \psi(z)}{dz^2}+q\varphi(z)\psi(z),
\end{equation}
where ${\cal E}_0$ 
is the minimal eigenvalue of the corresponding Hamiltonian with the explicit form given below.
This equation has to be complemented with 
the boundary conditions for $\psi(z)$,
\begin{equation}\label{eq:bcwavefun}
    \psi(0) = 0,
    \quad
    \psi(\infty)=0,
\end{equation}
since the dielectric is considered to be impenetrable for particles, see Sec.~\ref{sec:El_eqs}. The second condition in Eq.~\eqref{eq:bcwavefun} is necessary to fulfill the normalization condition~\eqref{eq:normwavefun}.  The equation for the electrostatic potential is the Poisson equation as before, see Eq.~\eqref{eq:Poisson1}, which can be expressed by means of Eq.~\eqref{eq: cond} in the way:
\begin{equation}\label{Poisson3}
    \frac{d^2 \varphi(z)}{dz^2} = -4\pi q\left(\frac{\varsigma}{\lambda_{T}^{3}} g_{{3}/{2}}\left(e^{-\beta q\varphi(z)}\right)+N_{0}(T) |\psi(z)|^{2}\right).
\end{equation}
We stress here that we can no longer reduce the order of differential equation, as it was done in the derivation of Eq.~\eqref{newreduced}.
Therefore, Eqs.~\eqref{normcond2},~\eqref{eq:grosspitaev}, and \eqref{Poisson3} represent a full set of the SCF equations that, taking into account Eqs.~\eqref{eq:normwavefun} and \eqref{eq:bcwavefun}, describes the system of charged bosons above the dielectric surface in the presence of a BEC. In general case, these equations can be solved only numerically.

The resulting equations require some comments, in particular, the Gross-Pitaevskii equation~\eqref{eq:grosspitaev}. The form of this equation is intuitive, but it can also be obtained in various ways, for example, from the requirement of the minimization of the functional~$R$,
\begin{equation}\label{func}
    R = \langle \tilde{\psi}|\hat{H}| \psi \rangle - {\cal E}_0 
    \langle\tilde{\psi}|\psi\rangle,
\end{equation}
where $\hat{H}$ is the Hamiltonian of the system
and ${\cal E}_0$ is a Lagrange multiplier ensuring the normalization of the wave function~$\psi$. Methods to derive the Gross-Pitaevskii equation can be found, e.g., in Refs.~\cite{Pethick2002,Pitaevskii2003}. 
For this reason, we do not include our method of derivation, which is based on variation of the functional~\eqref{func} and a synthesis of methods presented in Refs.~\cite{Slater1930,Gardiner2017}. 
We note only that Eq.~\eqref{eq:grosspitaev} can be derived schematically by employing the Heisenberg equation of motion for the field operators $\hat{\psi}(z,t)$ and $\hat{\psi}^{\dagger}(z,t)$,
\begin{equation}\label{eq: heisenberg}
    i\hbar\frac{\partial \hat{\psi}(z,t)}{\partial t} =
    \left[\hat{\psi}(z,t), 
    \hat{H} \left[ \hat{\psi}(z,t)\right]\right],
\end{equation}
where the Hamiltonian of the system is given by 
\begin{equation}\label{eq: hamiltonian}
    \hat{H} \left[ \hat{\psi}(z,t)\right]=
    \int_{0}^{\infty}dz
    \hat{\psi}^{\dagger}(z,t)
    \left(-\frac{\hbar^2}{2m}\frac{\partial^2}{\partial z^2}+q\varphi(z)\right)
    \hat{\psi}(z,t)
\end{equation}

Let us now evaluate the commutator on the right hand side of Eq.~\eqref{eq: heisenberg} and use the fact that for particles forming BEC the field operator $\hat{\psi}(z,t)$ can be replaced by a $c$-number, $\hat{\psi}(z,t)\to\psi(z,t)$. Since for stationary states 
    $\psi(z,t) = e^{i\mu t/{\hbar}}\psi(z),$
this yields Eq.~\eqref{eq:grosspitaev}, where instead of the quantity ${\cal E}_0$ 
the chemical potential $\mu$ is present. In other words, the Lagrange multiplier can be identified with the chemical potential $\mu$ (see also Refs.~\cite{Pethick2002,Pitaevskii2003}), so that ${\cal E}_0= \mu$.
We, however, cannot think so within the framework of our model. It is clear from Eqs.~\eqref{eq: cond}, \eqref{normcond2}, and \eqref{Poisson3} that within the developed approach it is necessary to set the chemical potential to zero below the critical temperature, $\mu(T\leq T_c) = 0$. 
Indeed, in the process of numerical solution of the SCF equations we obtain ${\cal E}_0 > 0$, i.e., $\mu > 0$. 
But for bosons described in terms of the Wigner function~\eqref{distr_fun} supplemented by the boundary conditions~\eqref{boundE}
\change{the chemical potential must be set to zero to yield positive occupation numbers of particles.
The given restriction for the chemical potential may lead to partially nonphysical results and we analyse these aspects in Subsec.~\ref{subsec:3C}.
}

On the one hand, our model keeps the possibility to employ the Wigner function meaning that the theory is quasiclassical in some sense. On the other hand, the Gross-Pitaevskii equation describes the BEC component of the system, which is a pure quantum phenomenon. 
In this context, the calculations show that as the Planck constant $\hbar\to 0$, the quantity ${\cal E}_0$ also vanishes. 
Thus, the chemical potential below the critical temperature is, at least, of the first order in $\hbar$, which agrees with semiclassical approximation for the distribution function~\eqref{distr_fun}, see also Eq.~\eqref{density}. It should be noted that the same situation is observed in a theoretical description of the trapped ultracold atomic gases with a BEC, \change{where one neglects the zero-point energy within the semiclassical approach,} see, e.g., Refs.~\cite{Pethick2002,Pitaevskii2003}.

Howbeit, Eqs.~\eqref{normcond2}, \eqref{eq:grosspitaev}, and \eqref{Poisson3} constitute a closed system of the SCF equations for the analysis of the system of charged bosons above the dielectric surface in the presence of a BEC. Solutions of these equations, in general case, can be accessed only through numerical algorithms. 
In the next section, we analyze the physical properties of the system under study by means of these procedures.

Note that the introduced approach has a necessary potential to be further modified to describe both condensate and thermal components on equal footing.
In general, one can express the system Hamiltonian as follows
\begin{equation}\label{eq:ham3d}
\hat{H} \left[ \hat{\psi}({\bf r},t)\right]=
    \int dV
    \hat{\psi}^{\dagger}_i({\bf r},t)
    \left(-\frac{\hbar^2}{2m}\nabla^2+q\varphi({\bf r})\right)
    \hat{\psi}_i({\bf r},t),
\end{equation}
where the summation over all eigenstates labelled by the index~$i$ is assumed.
Compared to the original ``hybrid'' approach, see Eq.~\eqref{eq: hamiltonian}, both the lowest eigenstate~$\psi_0$ (the condensate component below $T_c$) and excited states ($i>0$) can be evaluated numerically by setting appropriate boundary conditions in the corresponding spatial directions.
The choice of the optimal set of wave functions for the given geometry and temperature range represents then a separate problem, which can be studied and effectively solved with modern numerical algorithms.

In particular, by assuming a square box potential in the $xy$ plane as well as a separable wave function of the type $\psi({\bf r})=\sigma(x,y)\psi(z)$ and averaging over corresponding coordinates in transverse directions, we obtain the equation for the density~$n(z)$,
\begin{equation}\label{eq:nz-quant}
    n(z)= {\varsigma}{\lambda_{T}^{-2}}
    \sum_{i=0}^\infty |\psi_i(z)|^2\ln
    \left[
        \frac{1}{1-e^{-\beta(\varepsilon_i-\mu)}}
    \right].
\end{equation}
Note that the logarithmic dependence under the sum appears due to substitution of sums over quantum numbers associated with transverse degrees of freedom in $xy$ plane by corresponding integrals and evaluating the latter.
Obviously, the obtained equation describes the condensate and thermal components of the charged gas on equal footing, thus it becomes more accurate in comparison with Eq.~\eqref{eq: cond}. Therefore, Eqs.~\eqref{eq:Poisson1}, \eqref{eq:ham3d}, and \eqref{eq:nz-quant} form a closed set of equations describing the system beyond the semiclassical approximation.

\section{Results}\label{sec:results}
\subsection{Physical properties above $T_c$}
\begin{figure}[t]
  \includegraphics[width=\linewidth]{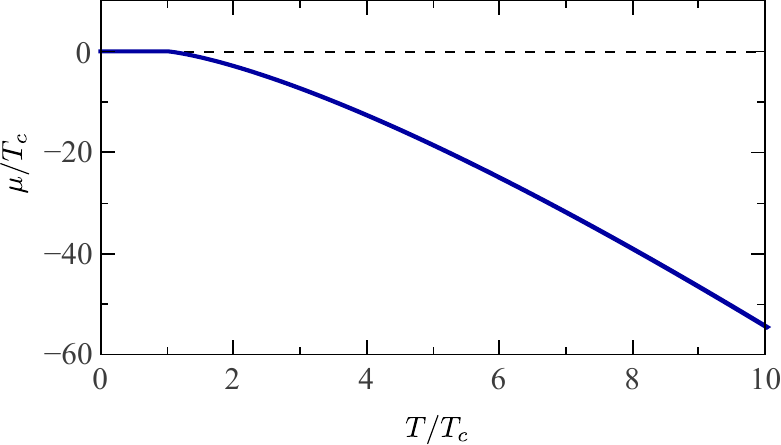}
  \caption{\label{fig:chempot}%
    Dependence of the chemical potential of the charged quasineutral gas above the dielectric surface on the temperature at the fixed number of particles $N_S$.}
\end{figure}

To analyze numerically the field and the charge distributions for the quasineutral system, we rewrite Eq.~\eqref{eq_of_state} in terms of the dimensionless variable $\tau=T/T_c$,
\begin{equation}
    g_{5/2}
    \{\exp[\mu(\tau)/\tau]\}
    =\zeta(5/2)\,\tau^{-5/2}.
\end{equation}
It is now obvious that the chemical potential depends only on the ratio $T/T_c$. Its universal behavior in units of $T_c$, which is determined according to Eq.~\eqref{eq:Tc-qneutr}, is shown in Fig.~\ref{fig:chempot} 

Now, let us analyze the field distribution above $T_c$ for the quasineutral system. To this end, we set $E_z(\infty)=0$ in Eq.~\eqref{newreduced} and express the latter as
\begin{equation}\label{EaboveTc}
    \frac{d\varphi}{dz}
    = 4\pi q N_S \tau^{5/4}
    \sqrt{ 
    g_{{5}/{2}}\left(e^{\beta\mu-\beta q\varphi(z)}\right)/\zeta(5/2)
    }.
\end{equation}

Introducing the dimensionless spatial variable~$\tilde{z}=z/L$ and the dimensionless potential~$\phi= q\varphi/T_c$, where
\begin{equation}\label{eq:Ldef}
    L=\left(
    \frac{\hbar^2}{8\pi N_S mq^2}
    \right)^{1/3},
\end{equation}
this can be rewritten as
\begin{equation}\label{EaboveTc2}
    \frac{d\phi}{d\tilde{z}}
    = a\,
    \tau^{5/4}
    \sqrt{ 
    g_{{5}/{2}}\left(e^{(\mu-\phi)/\tau}\right)/\zeta(5/2)
    },
\end{equation}
where $a={4\pi q^2 N_SL}/{T_c}$. In particular, for a gas consisting of the singly-ionized $^{40}$K atoms with
$N_S=10^{8}$~cm$^{-2}$ we obtain $L\approx0.3$~{nm} and $a\approx500$.

\begin{figure}[t]
  \includegraphics[width=\linewidth]{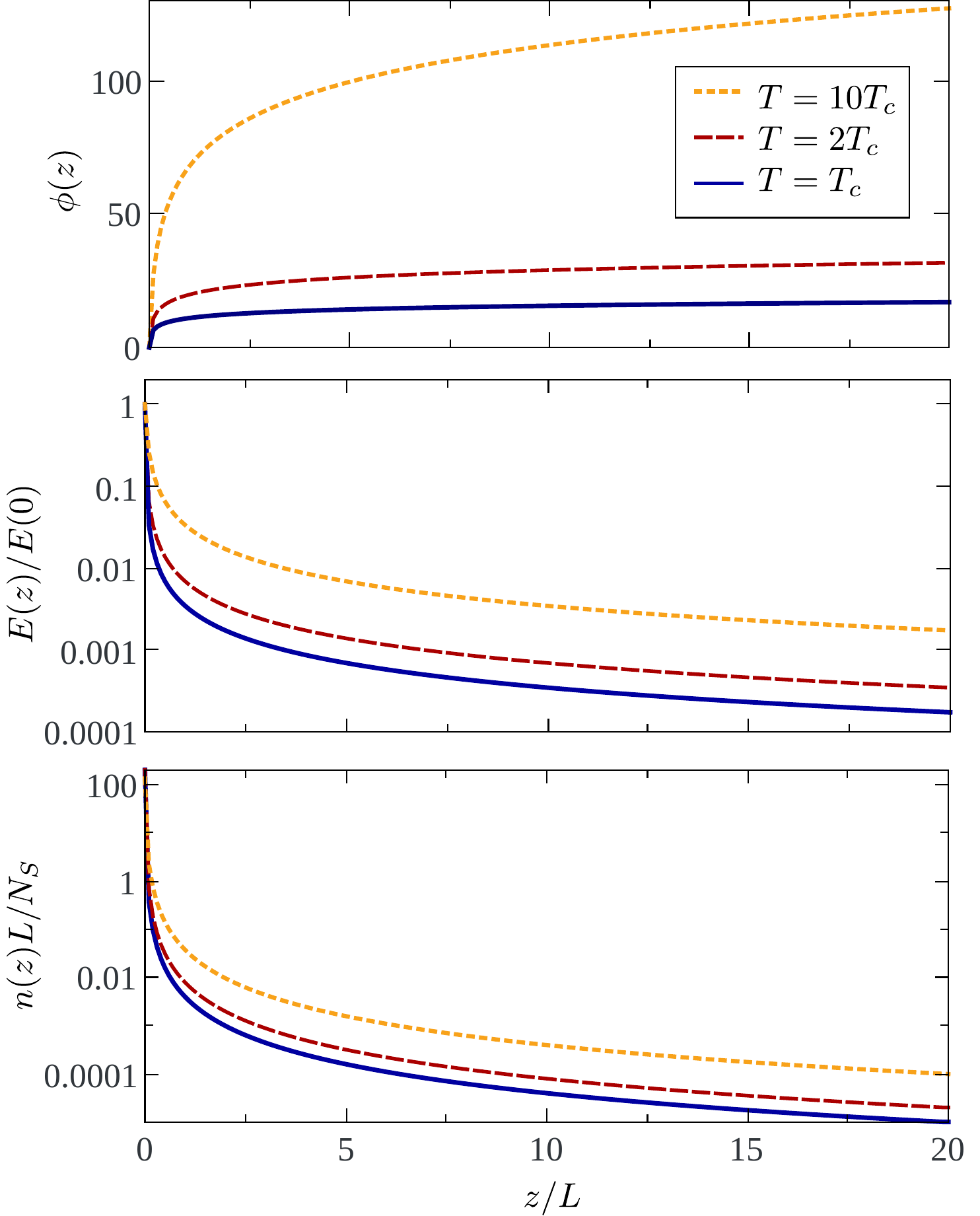}
  \caption{\label{fig:aboveTc}%
    Distributions of the electrostatic potential (upper panel), electric field (middle panel), and the density of charged particles (lower panel) on the distance from the dielectric surface at $a=500$.
    For sake of visibility, the logarithmic scale on certain vertical axes is used.}
\end{figure}

In Fig.~\ref{fig:aboveTc} we show the distributions of the electric field and the density of charged particles above the dielectric. Both quantities can be characterized as the rapidly decreasing functions on the given length scale~$L$. In particular, from the density distributions we conclude that for the charged gas consisting of singly-ionized atoms with $N_S\sim10^{8}$~cm$^{-2}$, the thermal cloud becomes almost completely confined in a narrow region of the order of a few nanometers above the dielectric surface.

\subsection{Physical properties below $T_c$}
Following the SCF formalism developed in Sec.~\ref{sec:formalism_below} and the notations introduced above, the Poisson equation~\eqref{Poisson3} can be written as
\begin{equation}\label{Poisson4}
    \frac{d^2\phi}{d\tilde{z}^2}
    =
    - \frac{a^2\tau^{3/2}}{2\zeta(5/2)}
    g_{{3}/{2}}
    \left( e^{-\phi(\tilde{z})/\tau}
    \right)
    -aN_0|\psi(\tilde{z})|^2/N_S.
\end{equation}
The contribution of the first term related to the thermal component grows with the temperature increase. As temperature reaches the critical value, $T=T_c$, Eq.~\eqref{Poisson4} can be transformed into Eq.~\eqref{EaboveTc2} by setting the condensate density to zero, $N_0=0$.

Below, we solve Eq.~\eqref{Poisson4} numerically by supplementing it with the minimization procedure for the energy functional governed by the Hamiltonian~\eqref{eq: hamiltonian}.
In dimensionless variables, this functional can be written as
\begin{equation}\label{eq:Efunc}
    {\cal E} \left[ \psi\right]/T_c=
    \int_{0}^{\infty}d\tilde{z}
    \left(a|\nabla\psi(\tilde{z},t)|^2+\phi(\tilde{z})|\psi(\tilde{z},t)|^2
    \right),
\end{equation}
where the corresponding minimal value~${\cal E}_0 = \lim_{t\to\infty } {\cal E}$ is determined through a dynamical evolution in time~$t$, i.e., the iterative numerical minimization procedure with the normalization condition~\eqref{eq:normwavefun}, see Ref.~\cite{Bao_2013} for more details.

Technically, below $T_c$ the SCF procedure is convenient to begin from the numerical solution of Eq.~\eqref{Poisson4} with no condensate, $N^{(0)}_0=0$. The calculated in this way potential generated by the thermal component $\phi_T^{(1)}(\tilde{z})$ with imposing the boundary conditions for a quasineutral system allows to obtain $N^{(1)}_T$ and thus $N^{(1)}_0$.
This, in turn, provides an estimate for the contribution of the condensate component $\phi_0^{(1)}(\tilde{z})$ by means of the Poisson equation with the corresponding right-hand side.
The function $\phi_0^{(1)}(\tilde{z})$ is updated during the iterative procedure of minimizing the energy functional~\eqref{eq:Efunc}, since the solution of the corresponding Poisson equation depends on the wave function~$\psi(\tilde{z})$.
The updated functions $\phi_0^{(1)}(\tilde{z})$ and $\psi^{(1)}(\tilde{z})$ are subsequently put back into the original Poisson equation~\eqref{Poisson4}, which is solved then with a full account of both condensate and thermal components. The condensate fraction $N^{(2)}_0/N_S$ can be viewed in this step as an adjustable parameter to fulfill boundary conditions for the electric field in a quasineutral system. The remaining solution of the Poisson equation for the condensate component and the energy minimization are performed with the new value of the condensate density. After the new minimum is reached, the whole SCF procedure is repeated till the final convergence.

\begin{figure}[t]
  \includegraphics[width=\linewidth]{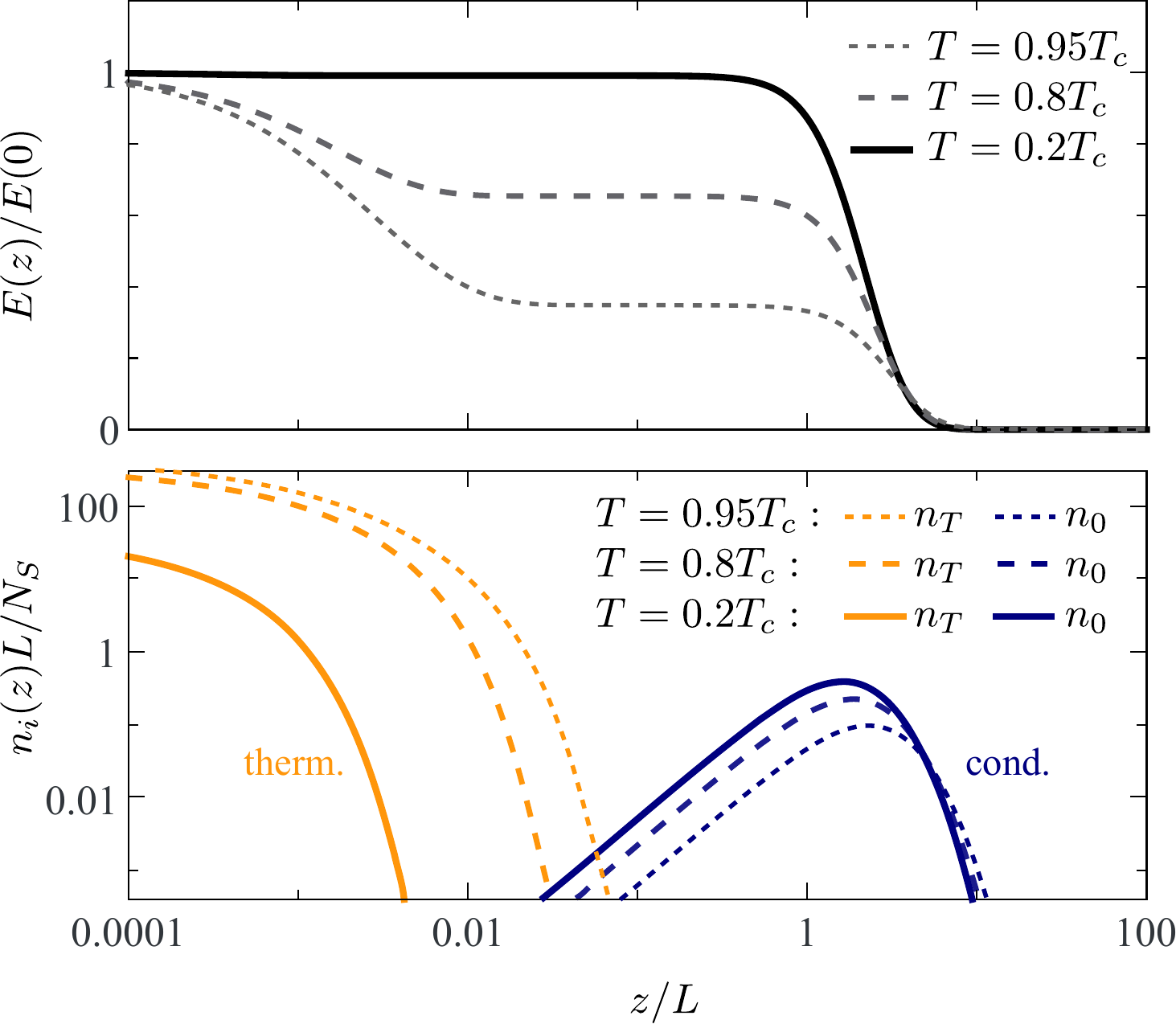}
  \caption{\label{fig:belowTc}%
    Distributions of the electric field (upper panel) and the partial densities of charged particles (lower panel) on the distance from the dielectric surface in the presence of a BEC at $a=500$.
    For sake of visibility, the logarithmic scale on certain axes is used.
    }
\end{figure}
The resulting field and density distributions below $T_c$ are shown in Fig.~\ref{fig:belowTc}. 
For compactness, we do not show the distributions of the potential, since the presence of a BEC does not change the potential profiles qualitatively, see Fig.~\ref{fig:aboveTc} as a reference (the main effect is an increase of absolute values of $\phi$ with the temperature decrease).
In contrast, the field and the density distributions show significant qualitative changes in comparison with their behavior above $T_c$.
In particular, in the field distributions one can notice a step-like descent that corresponds to the presence of two spatially-separated charged regions in the system under study below $T_c$.
From the corresponding density profiles shown in Fig.~\ref{fig:belowTc} it becomes clear that this behavior originates from a strong confinement of the thermal component at the surface, while the condensate component remains substantially separated with the corresponding (almost $T$-independent) peak position at $z\approx 2L$.

\begin{figure}[t]
  \includegraphics[width=\linewidth]{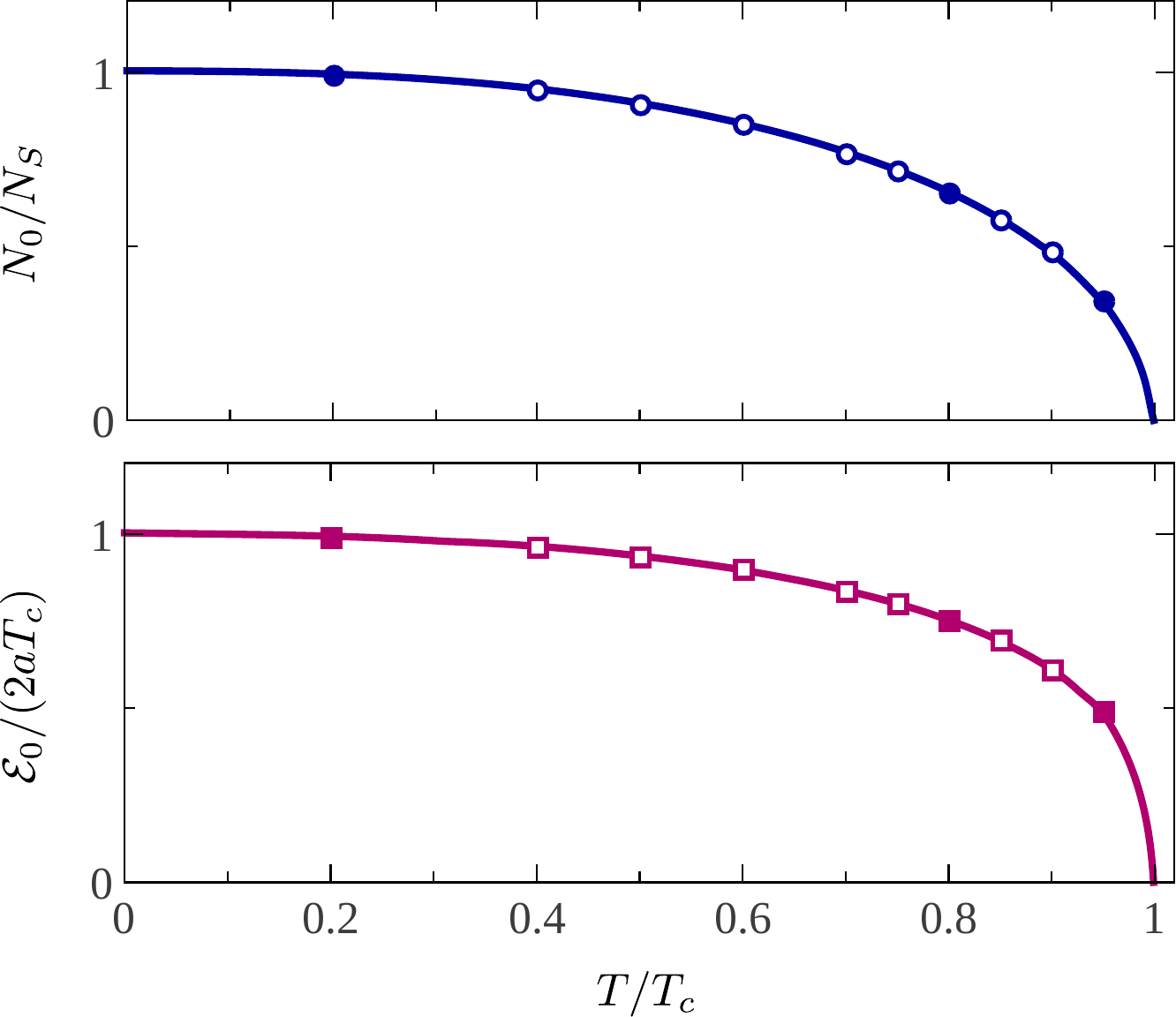}
  \caption{\label{fig:belowTc2}%
    Dependencies of the fraction of particles in the BEC state (upper panel) and the corresponding energy at $a=500$.
    The symbols refer to the results of numerical calculations and lines serve as a guide to the eye. The filled symbols at $T=0.2T_c$, $T=0.8T_c$, and $T=0.95T_c$ correspond to the distributions shown in Fig.~\ref{fig:belowTc}.
    }
\end{figure}

In Fig.~\ref{fig:belowTc2} we also show the temperature dependencies of the number of particles in the BEC state \eqref{normcond2} and the lowest eigenvalue of the Hamiltonian~\eqref{eq: hamiltonian} entering the Gross-Pitaevskii equation~\eqref{eq:grosspitaev}. 
These are the monotonously decreasing functions of temperature vanishing at $T=T_c$, where the applicability of the approach on its own becomes limited.

\subsection{Description beyond the semiclassical approximation}\label{subsec:3C}

We must note that the density distributions shown in Figs.~\ref{fig:aboveTc} and~\ref{fig:belowTc} contain a nonphysical behavior in the limit $z\to0$.
This is a consequence of the employed semiclassical approach, which treats all particles as point-like objects. In reality, the charge density must decrease and vanish at $z=0$ on the length scales of the order of the Bohr radius. To obtain more accurate dependencies in this limit, it is necessary to turn to a quantum-mechanical description in terms of wave functions for both the condensate and thermal components.

\begin{figure}[t]
  \includegraphics[width=\linewidth]{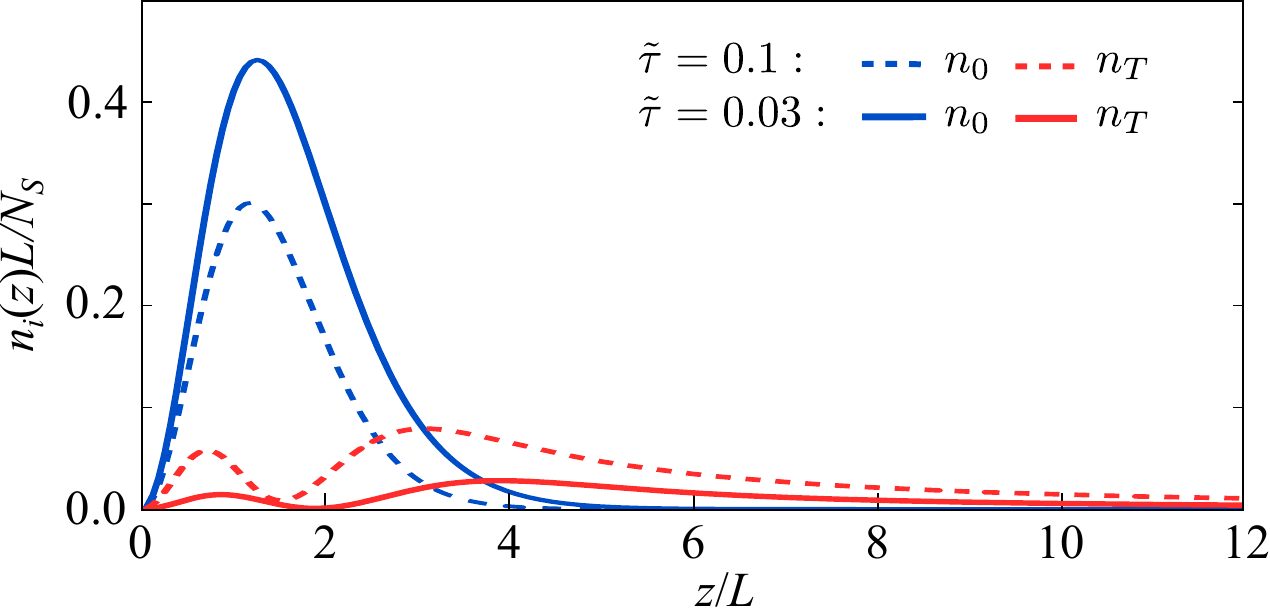}
  \caption{\label{fig:quant}%
    Distributions of the densities of components obtained within the quantum-mechanical approach at $N_S=10^8$~cm$^{-2}$ and two different temperatures $\tilde{\tau}$, where $n_T$ corresponds to the sum over excited states with $i>0$ in Eq.~\eqref{eq:nz-quant}.
    }
\end{figure}

Below, we provide a characteristic example, where this issue is effectively solved by employing the quantum-mechanical description on the basis of  
Eqs.~\eqref{eq:Poisson1}, \eqref{eq:ham3d}, and \eqref{eq:nz-quant}. We study the effect for the case when the ground- and exited-state components have comparable contributions to the equation determining the chemical potential~$\tilde{\mu}$,
\begin{equation}
        \tilde{\tau}
        \sum_{i=0}^\infty
    \ln
    \left[
        \frac{1}{1-e^{-(\varepsilon_i-\tilde{\mu})
        /4\pi\tilde{\tau}}}
    \right]
    = \chi_S,
\end{equation}
where the dimensionless parameters $\tilde{\tau}=mTL^2/2\pi\hbar^2$ and $\chi_S=N_S L^2/\varsigma$ related to the temperature and the number of particles, respectively.
To this end, we take $\chi_S=10^{-8}$ (this corresponds to $N_S=10^8$~cm$^{-3}$, thus $a\approx500$ as above) and vary the effective temperature in the limits $\tilde{\tau}\in[0.01,0.1]$. Within this temperature range and parameters of the system, we determined that it is sufficient to account for up to 80 excited states in numerical analysis.
The resulting distributions of the condensate and thermal components are shown in Fig.~\ref{fig:quant}. 
As one can see from the figure, the density of the thermal component $n_T$ vanishes with $z\to 0$, i.e., shows qualitatively correct physical behavior.
Note also that within the quantum-mechanical treatment, a significant weight of the thermal component becomes distributed further away from the dielectric surface.


\section{Conclusions and Outlook}\label{sec:conclusion}
We theoretically studied key aspects of Bose-Einstein condensation in a gas consisting of indistinguishable charged bosons above a flat surface of dielectric. 
It is shown that stability of the system can be realized only in the presence of an additional electric field holding charges at the dielectric surface.
The chosen research direction, in particular, is motivated by a class of problems thematically connected to the so-called Wigner crystallization phenomenon.
This periodic spatial arrangement can be realized in systems consisting of electrons above the surface of liquid helium placed on a dielectric substrate.
The aspects of the experimental realization of similar systems in Bose gases are left beyond the scope of the current study.

To describe main physical properties and phenomena in a charged Bose gas, we introduced the SCF approach, which combines a quasiclassical description employing the Wigner distribution functions for the thermal component and a quantum-mechanical approach involving the Gross-Pitaevskii equation.
In the framework of the developed theory we provided a relatively complete description of the state of the system both above and below the phase transition to the BEC state.
We determined the conditions for this phase transition with the corresponding critical temperature as a function of the number of bosons in a gas and the intensity of the external field.
We analyzed how the BEC fraction affects the main physical properties, such as the field and charge distributions above the dielectric surface.
In particular, it is shown that there is a significant spatial separation of the thermal and condensate components dictated by the tendency of fractions to be localized on different distances from the surface. 

It should be mentioned that many aspects of the potential experimental realizations are left beyond the scope of the current study. 
We assume that the novelty and uniqueness of the obtained theoretical results will stimulate further studies oriented on the realization of these systems under experimental conditions.
In particular, one can draw a formal analogy to the systems consisting of electrons above the films of liquid helium, where the periodic structures (two-dimensional Wigner crystals) are observed. Here, in contrast to Fermi gases, an additional degree of freedom associated with the BEC formation opens new prospects to predict and study ``hybrid'' phenomena, in particular, a coexistence of two phases in Bose gases: a BEC and periodic ``crystalline'' structures.

\begin{acknowledgments}
The authors acknowledge support by the Ministry of Education and Science of Ukraine, Research Grant No.~0120U102252 and the National Research Foundation of Ukraine, Grant No.~0120U104963.
\end{acknowledgments}


\bibliography{bec_charged}
\end{document}